\documentstyle[12pt,aaspp4]{article}
\begin{document}
\tightenlines
%
%
\newcommand{\ea}{{et~al.}}
\newcommand{\IUE}{{\it IUE}}
\newcommand{\dmod}{$(m - M)_{0}$}
\newcommand{\logg}{$\log g$}
\newcommand{\logl}{$\log L$}
\newcommand{\lya}{\mbox{Ly$\alpha$}}
\newcommand{\lsun}{L$_{\sun}$}
\newcommand{\msun}{M$_{\sun}$}
\newcommand{\mv}{{M}$_{V}$ }
\newcommand{\rsun}{R$_{\sun}$}
\newcommand{\teff}{T$_{\rm eff}$}
\newcommand{\vsini}{$v \sin i$}
\newcommand{\rd}{Di\thinspace Stefano}
\newcommand{\re}{Einstein radius}
\newcommand{\er}{Einstein ring}
\newcommand{\ml}{microlensing}
\newcommand{\lr}{lensing region}
\newcommand{\pl}{planet}
\newcommand{\mage}{magnification}
\newcommand{\lc}{light curve}
\newcommand{\ev}{event}
\newcommand{\ec}{encounter}
\newcommand{\pop}{population}
\newcommand{\mo}{monitoring}
\newcommand{\bi}{binary}
\newcommand{\bis}{binaries}
\newcommand{\ob}{observation}
\newcommand{\ps}{planetary system}
\newcommand{\sy}{system}
\newcommand{\pr}{program}
\newcommand{\res}{resonant}
\newcommand{\ov}{overlap}
\newcommand{\cs}{central star}
\newcommand{\inm}{innermost planet}
\newcommand{\rpe}{repeating event}
\newcommand{\rp}{repeat}
\newcommand{\is}{isolated}
\newcommand{\dbn}{distribution}
\newcommand{\shdn}{short-duration}
\newcommand{\evd}{evidence}
\newcommand{\zres}{zone for resonant lensing}
\newcommand{\sep}{separation}
\newcommand{\dtn}{detection}
\newcommand{\bl}{blending} 
\newcommand{\mt}{monitoring teams} 
\newcommand{\fut}{follow-up teams} 
\newcommand{\ft}{follow-up teams} 
\newcommand{\fss}{finite-source-size} 
\newcommand{\fsse}{finite-source-size effects} 
\newcommand{\cc}{caustic crossing}
\newcommand{\aicn}{and its companion (Di\thinspace Stefano \& Scalzo 1998)}
\newcommand{\Asun}{A$_{\sun}$}
\newcommand{\Ajup}{A$_J$}
\newcommand{\Anep}{A$_N$}
\newcommand{\Nsun}{N$_{\sun}$}
\newcommand{\goesto}{\longrightarrow}
\def\mr{multiple repetitions}
\def\op{orbital plane} \def\lp{lens plane}
\def\wo{wide orbit}
\def\w-o{wide-orbit}
\def\otn{orientation} \def\vy{velocity} \def\vt{$v_t$}
\def\smn{simulation}
\def\em{Earth-mass}
\def\jm{Jupiter-mass}


\def\stacksymbols #1#2#3#4{\def\theguybelow{#2}
    \def\verticalposition{\lower#3pt}
    \def\spacingwithinsymbol{\baselineskip0pt\lineskip#4pt}
    \mathrel{\mathpalette\intermediary#1}}
\def\intermediary#1#2{\verticalposition\vbox{\spacingwithinsymbol
      \everycr={}\tabskip0pt
      \halign{$\mathsurround0pt#1\hfil##\hfil$\crcr#2\crcr
               \theguybelow\crcr}}}

\def\lapproxeq{\stacksymbols{<}{\sim}{2.5}{.2}}
\def\gapproxeq{\stacksymbols{>}{\sim}{3}{.5}}
\def\du{duration} \def\dca{distance of closest approach}
\def\stl{stellar-lens}
\vskip -.9 true in
\title{
A New Channel for the Detection of Planetary Systems Through Microlensing: 
I. Isolated Events Due to Planet Lenses}

\author{Rosanne \rd\altaffilmark{1},
Richard A. Scalzo\altaffilmark{2}}

\altaffiltext{1}{
Harvard-Smithsonian Center for Astrophysics,
60 Garden St., Cambridge, MA 02138; e-mail:  rdistefano@cfa.harvard.edu}

\altaffiltext{2}{Department of Physics, University of Chicago,
Chicago, IL 60637; e-mail:  rscalzo@rainbow.uchicago.edu}

\vspace{-0.15in}

\begin{abstract}

\vspace{-0.15in}
We propose and evaluate the feasibility of a new strategy to search for \pl s
via \ml\ \ob s.
This new strategy is designed to detect \pl s
in ``wide" orbits, i.e., with orbital separation, $a$,
greater than $\sim 1.5 R_E$.  
Planets in \wo s may provide the dominant channel for the  
discovery of \pl s via \ml, particularly low-mass (e.g., \em ) \pl s.
This paper concentrates 
on \ev s in which a single \pl\ serves as a lens, leading to
an isolated \ev\ of \shdn . 
We point out that 
a distribution of events due to lensing
by stars with wide-orbit planets
is necessarily accompanied by a distribution
of shorter-duration events.  The fraction of events in the latter
distribution is proportional to the 
average value of $\sqrt{q}$, where $q$ is the ratio between
\pl\ and stellar masses. The position of
the peak or peaks also provides a measure of the mass ratios typical of
planetary systems. We study detection strategies that
can optimize our ability to discover 
isolated \shdn\ \ev s due to lensing by \pl s, and find that monitoring
employing sensitive photometry is particularly useful. 
If \ps s similar to our own are common, even 
modest changes in detection strategy should lead to the discovery  
a few  isolated \ev s of
\shdn\ every year.  
We therefore also address the issue of
the contamination
due to stellar populations of any \ml\ signal due to low-mass MACHOs.
We describe how,
even for isolated \ev s of \shdn , it will be possible to
test the hypothesis that the lens was a planet instead of a low-mass
MACHO, if the \cs\ of the \ps\ contributes a measurable fraction
of the baseline flux.
\end{abstract}
\vskip -.2 true in

\section{Introduction}

When planets are discovered through monitoring programs similar to those 
presently being carried out, the distance to the lens will typically be on the
order of kiloparsecs.
Although some follow-up may be possible (\rd\ \& Scalzo 1997,  1998, 
\rd\ 1998a,b), 
imaging, such as that planned
to probe possible companions to nearby stars (Stahl \& Sandler
1995; Bender \& Stebbins 1996; Labeyrie 1996; Angel \& Woolf 1997)
will not be possible in the near
future.
Nevertheless, planets discovered as microlenses can play an important role in
developing our understanding of planetary systems in our Galaxy and beyond.
The reason for this is precisely because microlensing observations probe vast
volumes of the Milky Way and other galaxies, such as the Magellanic Clouds and M31.  Thus,
microlensing provides a unique window for studying the statistics of planetary
systems (numbers and properties) and their dependence on the local stellar
environment.  Microlensing searches complement velocity-based searches in
another way as well.  Particularly for the wide systems studied here for the
first time, the microlensing searches can be effective for low-mass planets 
orbiting at low speeds; such \pl s  cannot readily be detected via radial velocity
methods.

The framework for the work to date on discovering planets through \ml\
was established by Mao \& Paczy\'nski (1991) and Gould \& Loeb (1992).
These authors found that when the separation, $a$, between the star
and \pl\ is close to the Einstein radius, $R_E,$ of the star 
($0.8 R_E  \lapproxeq a \lapproxeq 1.5 R_E$),
a significant
fraction of \ev s ($\sim 5-20\%$) in which the star serves as a lens
would be perturbed in a detectable way. 
This has been referred to as ``resonant" lensing,
both because the separation must be close to $R_E$ in order for
the signal to be detectable, and because the signal itself is
sharp and distinctive. 
The detection strategy is to monitor light from an
ongoing event at frequent intervals in order to observe a short-lived
perturbation.
With the analysis of more than $200$ \ml\ \ev s reported to date,
it is not clear whether any planetary-lens \ev s have been discovered.
\footnote{
One candidate for a resonant planetary lensing event has been suggested
(Bennett {\it et al.} 1996a).
There are, however, two reasons to be cautious
about the interpretation of this event.  First, the mass ratio derived from
the binary-lens fit is $\sim 0.043$, which is large enough to be consistent
with lensing by a binary stellar system.  Second, the degeneracy of the
physical solution has not yet been worked out.  The degeneracy may be of two
types:  (a)\ Other binary solutions with values of $q$ differing from the one
for this fit by as much as an order of magnitude may prove to be equally good
fits (Di Stefano \& Perna 1996); this needs to be systematically checked.
(b)\ Other physical effects may prove to be important, making the determination
of the system's true parameters even more difficult. These effects include 
finite source size
effects (Witt \& Mao 1994; Witt 1995), finite lens-size effects
(Bromley 1996), and 
blending (\rd\ \& Esin 1995, Kamionkowski 1995).
(In fact blending has been  
a feature of every other binary-lens event; Udalski {\it et al.} 1994,
 Alard, Mao, and Guibert 1995, Alcock {\it et al.} 1997a,
Bennett {\it et al.} 1996b.)}

The purpose of this paper \aicn\ is to develop a complementary framework for the
discovery of \pl s by \ml . The basic idea is that planets located more than
$\sim 1.5 R_E$ from the central star can give rise to 
characteristic \ml\ signatures that reveal
evidence of their presence.  
The advantages of searching for \pl s in \wo s are that there are likely to be   
be several such \pl s for every \pl\ that can give rise to a resonant
\ev , and the probability of detection is not decreased in the same way
\footnote{The influence of finite source size on the detection of \pl s in the \zres\
depends on the location of the \pl\ within the zone. The net effect has been calculated
in most detail by Bennett \& Rhie (1996). See \rd\ \& Scalzo 1997, 1998 for a
specific comparison
of \fsse\ on the detection of \pl s in \wo s with the detection of \pl s in the \zres .      
}, and
can even be increased, by \fsse .

We will say that planetary orbits in which the
separation between the star and planet is larger than roughly 1.5 $R_E$ are ``wide".
(See \S 2.3.)
There are two types of \ev s that provide evidence for the presence of
\pl s in \wo s. The first are isolated \shdn\ \ev s due to the passage of the 
track of a  distant star through the \er\ (or \lr ) of an intervening
\pl . The second are \rpe s,
in which the track of an individual lensed source passes through the \lr s of
more than one mass in the \ps . Among repeating \ev s,   
the probability is largest that the
track of the source will pass through the \lr s of
the \cs\ and one \pl , but it can also pass through the rings 
of several \pl s.
This paper 
focuses
on \is\ \ev s; the companion paper (\rd\ \& Scalzo 1998)
concentrates on \rpe s.

\S 2 lays the foundation for important parts of both papers.
Although comprehensive introductions to \ml\ can be found elsewhere (see, e.g.,
Paczyn\'ski 1996), \S 2  
includes a brief general introduction to \ml , as well as some material
specific to the case of lensing by \pl s in wide orbits. 
Of particular importance are (1) the notion of the width of the \lr , 
and (2) the role played by \fsse\ in the \dtn\ of \pl s in \wo s.
In \S 3 we
address the question of whether \pl s in \wo s are likely to
exist and to serve as lenses. Event rates, and detectability issues are the
subjects of \S 4.
In \S 5 we discuss the detection strategies that
can optimize the discovery of isolated \shdn\ \ev s and 
test the hypothesis that they are due to
lensing by \pl s;
we focus on methods that involve modifications of
existing observing programs.
We sketch our conclusions in \S 6.

\section{Technical Introduction}

\subsection{Microlensing}

Consider light traveling from a star located a distance $D_S$ from us.
If there is an intervening mass, $M,$ located a distance $D_L$ from us,
gravity will deflect the light from the path it would
otherwise have taken.
The deflection leads to multiple and generally distorted images of the source.
When $D_S$ is on the order of tens of kpc and the lens is a solar-mass star, 
the maximum  
image separations are on the order of milliarcseconds,
and are generally not resolvable.
Changes in the amount of light received can be measured however, and these
are also associated with lensing. The changes in flux are due to the
motion of the source relative to the lens along the direction transverse to
our line of sight. Let $u$ represent the instantaneous distance between the
source and the lens, as projected onto the lens plane. The instantaneous
value of the magnification, $A,$ is

\begin{equation} 
A={{u^2+2}\over{u \sqrt{u^2+4}}}.    
\end{equation} 

\noindent 
The magnification itself is due to the fact that, although the
surface brightness of the source is not affected by gravitational lensing,
the total surface area of the images does change as the projected position of the 
source moves closer to and farther away from the lens. In the above equation, $u$
is measured in units of the Einstein radius, $R_E$.

\begin{equation}
R_E=\Bigg[{{4\, M\, G\, D_S\,\, x\, (1-x)}\over{c^2}}\Bigg]^{{1}\over{2}}, 
\end{equation}

\noindent where $x=D_L/D_S$. For $u=1,$ the magnification is equal to $1.34.$  
This magnification is measurable; indeed, the Einstein radius
is often used as a measure of the width of the lensing region,
For a solar-mass lens, with $D_S= 10$ kpc and $x= 0.9,$ $R_E=2.7$ au.

In this paper and its companion (\rd\ \& Scalzo 1998), 
we will often set $D_S$ to 10 kpc and $x$ to 0.9. This
roughly corresponds to lensing by a star or \ps\ in the Galactic Bulge
of light from a more distant Bulge star.
\footnote{Smaller distances to Bulge stars
may apply to a large fraction of potential Bulge source stars. In these
cases, $x$ will typically be smaller, and the size of stellar and
planetary Einstein rings will not differ
substantially from the values we compute with $D_S = 10$ kpc and $x=0.9$.
}
 Because the greatest number of
\ev s have been discovered along the direction to the Bulge, and because
many of these may in fact be \ev s in which a Bulge star lenses light from a
more distant Bulge  star, these choices of the
parameters may be the ones most relevant for the near-term discovery of \pl s
via \ml .
The effects of other parameter choices
can be gauged from Eq. 2. In general, making $D_S$ larger, or choosing $x$
closer to $0.5$ (the value that maximizes the size of the \er\ for given choices
of $M$ and $D_S$), will increase the size of the Einstein radius,
and will require \pl s to be located farther from the \cs\ in order to
be in ``wide orbits" (\S 2.3). 

A measure of the \ev\ \du\ is given by 
$(2\ R_E)/v_T,$ where $v_T$ is the 
transverse velocity between source and lens. (See, e.g.,  Figure 6.)  
If $v_T= 100$ km/s, the event duration would be $\sim 94$ days,
if the lens is of 
solar mass. 
If the lens is a Jupiter-mass (Earth-mass) \pl , then the 
\ev\ duration would be $\sim 3$ days ($4$ hours).   
Clearly the short duration of \ev s in which low-mass objects serve as lenses
poses a challenge to the detection of \pl s via microlensing.  
We note, however, that when \fsse\ play a role, the relevant crossing time is
more closely related to the time taken for the track of the source to
move through a distance equal to the size of the source. If $R_S = 10\, R_\odot,$
and $v_T=100$ km/s, $R_S/v_T\sim 19$ hours. 
\subsection{The Standard Strategy for Microlensing Detections of Planets}

The strategy for \pl\ discovery suggested by 
Mao \& Paczy\'nski (1991) and Gould \& Loeb (1992) relies on observing 
an \ev\ due to lensing by the \ps 's  star, and detecting perturbations
from the form given by Eq (1). Although the time scale for the stellar-lens
\ev\ may be on the order of months, the perturbations
due to \pl s are short-lived,
lasting for times on the order of hours.
The perturbations are due to the passage of the 
track of the source through or very 
near to the caustic structure associated with the presence of the \pl .
If the lensed source is point-like, then there is an infinite
jump in the \mage\ as it passes through a caustic curve. Finite-source-size
effects moderate the magnitude and characteristics of the change. 
Caustics are a feature associated with lens multiplicity. The case
of a \pl\ orbiting a star can be thought of as a binary system with an
extreme mass ratio; generally we expect $q_p = m_{p}/m_{\ast}$ to be
less than 
$10^{-2}$ or $10^{-3}$.   
Because the mass of the \pl\ is small,
the caustic structures around the stellar position are small; this is why the 
perturbations of the stellar-lens \lc\ tend to be short-lived.

We will use the phrase ``\zres" to refer to the region 
($a_c < a < a_w$) in which
caustic crossing \ev s are most likely to occur. 
The spatial extent of the \zres\ was discussed by
Paczy\'nski (1996) and has been  studied most
recently by Wambsganss (1997). Our results
on detecting \pl s in \wo s do not rely heavily on
the exact position of its inner boundary,  $a_w$.

As presently implemented, the search for resonant planets is carried out in
two steps.  First, the monitoring teams (e.g., EROS, MACHO, OGLE)
identify \ml\ \ev\
candidates.  They do this by monitoring 
the flux from ${\cal O}(10^7$) stars regularly.
Some of the monitored fields are not visited every night; others
are the
targets of regular nightly monitoring and may occasionally be re-visited
even within a single night.
If the monitoring teams discover that an otherwise non-variable star's flux
has increased significantly above baseline in several consecutive
measurements, they typically issue an alert so that other observers can
monitor the star more frequently. Follow-up teams have been formed to take
systematic advantage of this opportunity
(Udalski et al. 1994; Pratt et al. 1995; Albrow et al. 1996).
Under
favorable conditions, the follow-up teams carry out hourly
monitoring with good photometry ($\sim 1\%$).

Although no
promising candidates for \ev s displaying clear signs of a \pl\ lens
have yet emerged from the data sets, there may be several reasons for this.
For example, the detection efficiency may be low enough that a larger number of events
need to be monitored in order to have a good chance of detecting a resonant
\ev . It might also be that 
planets with the appropriate distance from the central star are not
common.  
In addition, finite-source-size effects can wash out the signature
of the short-time-scale perturbations characteristic of planetary mass ratios,
decreasing the number of detectable \ev s (Bennett \& Rhie 1996).

\subsection{Planets in Wide Orbits}

The idea that the term ``wide orbit" is meant to capture is that,
as the orbital separation between the \pl\ and the \cs\ increases,
\cc\ \ev s
become less likely, while \ev s in which a \pl\ acts as an independent
lens become more common. These latter are characteristic of planet-lens \ev s,
when the \pl\ is in a \wo . 
If the track of the source passes through the Einstein ring of a 
single \pl , we may discover an isolated \ev\ of \shdn .
If the track of the lensed star passes through the  
Einstein rings of several masses in the \ps , then we may hope to 
discover a repeating
\ev . Although we have used the phrase ``Einstein ring", we note that,   
if the monitoring teams employ sensitive photometry, they 
can detect \ev s in which the distance of closest approach to the lens
is larger than an Einstein radius. It is therefore useful to 
introduce the notion of the ``\lr": 
the region around the lens in the lens plane
through which the source track must pass in order for a {\it detectable} \ev\
to occur.

We will say that a 
planet is in a ``wide" orbit if its distance from the central
star is large enough that the isomagnification contour associated with
$A = 1.34$ is comprised of two separated, closed curves, one centered near the
star and the other centered near the planet.  For mass ratios ranging from
$10^{-5}$ to $10^{-3}$, the critical orbital radius, $a_w$, beyond which an
orbit becomes wide is roughly equal to $1.5 R_E$.  
The choice of boundary, $a_w,$ is of course somewhat arbitrary.
Were the exact value of $a_w$ important, then its dependence on
the mass ratio would need to be taken into account. In fact, however,
its exact value makes little difference to results on wide-orbit \pl s;
these would be roughly the same had we chosen $a_w = 1.8\, R_E,$ 
or $a_w = 2.0\, R_E$,
The effect of increasing orbital separation
near the boundaries of the \zres\ has been studied in detail by 
Wambsganss (1997), and readers are referred there for a detailed discussion
and instructive graphics.      

\subsection{Width of the Lensing Region}

Consider a lens of mass $m_i$. 
As mentioned above, the lensing region associated with this mass
 is that region in the lens
plane through which the track of the source must pass in order for an
{\it observable} \ev\ to occur. If the region for lensing by a particular star
or \pl\ is a disk, as it very nearly is when the \pl s are in \wo s,
the width, $w_i,$ of the lensing region can be 
taken to be the radius of
the lensing disk. Because the size of the \lr\ relative to the \er\
is important, it is convenient to define 
\begin{equation}
n_i=w_i/R_{E,i}.  
\end{equation}  
$n_i$ is the maximum impact parameter for an {\it observable} \ev ,
expressed in units of the Einstein radius. Note that $n_i$ is not
generally an integer.   
Large values of $w_i$ facilitate \pl\ detection in two ways. First, the \ev\
probability is proportional to $w_i$. Second, the larger the value of $w_i,$
the longer typical \ev s will 
be observable, and the better our chances of \ev\
detection.

The value of $w_i$ can be made larger in two ways. The first is
simply to increase the photometric sensitivity of the \mo\ \pr s.
Let $A_{min}$ represent the minimum value of the peak \mage\ for which
an  \ev\ can be reliably detected. If $A_{min}=[1.58, 1.34, 1.06, 1.02],$
then $n = n_i = [0.76, 1, 2, 3].$
\footnote{Note that when \fsse\ are unimportant, $n_i$ has the same value, $n$,
for all lenses. Since, however, the role played by \fsse\
is specific to the lens, the value of $n_i$ {\it can} be lens-specific.}
The general formula for $n$ in terms of $A_{min},$ is
\begin{equation}
n=\sqrt2\, 
\Bigg[ \Big(1 + {{1}\over{A_{min}^2-1}}\Big)^{{1}\over{2}} - 1 \Bigg]^{{1}\over{2}}.
\end{equation}

The MACHO team has been driven to use
a higher value of $A_{min}$ ($1.58$) in order to eliminate the possibility
that some forms of stellar variability may mistakenly be identified as \ml\
(see, e.g., Purdue {\it et al.} 1995).
 As the star fields studied by the observing teams are monitored over   
longer times, however, unrecognizable stellar variability will become
less of a potential problem, and the \mo\ teams should be able to lower their
values for $A_{min}.$ The photometric sensitivity of the follow-up teams is
such that, were they to attempt to identify new \ev s in the fields where they
are \mo\ ongoing \ev s,  $A_{min} = 1.06$ ($w_i = 2\, R_E$) could be
achieved. 

The second way in which $w_i$ can be increased is if the size of the 
lensed star, as projected onto the lens plane, is 
comparable to or slightly larger
than the size of the Einstein ring. 
Although, finite source size tends to decrease the value
of the peak \mage , it can also increase the size of the \lr ,
since some parts of the source may be near enough to the lens position to
experience significant \mage , even when the center of the source
is farther away than a point source would have to be in order to experience
the same overall enhancement of flux.    

\subsection{Event Duration}

Our primary interest is in the detection of isolated \ev s of \shdn.
Thus, our definition of \ev\ \du\ must be keyed to the amount 
of time during which there is a deviation from the baseline flux
that would be discernible, given a particular level of
photometric sensitivity. If \ev s can be reliably detected when the 
peak \mage\ is $A_{min},$ then their start can certainly be
detected before the \mage\ is $A_{min},$ and they can continue to
be monitored after the \mage\ has fallen below $A_{min}.$ In order
to be conservative about the possibility of \ev\ discovery and
detection, we will say that an \ev\ starts when the \mage\
rises above $A_{min}$, and ends when the \mage\ falls 
below $A_{min}.$ Thus, the value of $A_{min}$ 
determines the width, $w_i,$ of the \lr . $w_i$ in turn
plays a key role in setting the time, $\tau_i,$ during which an
\ev\ is observable.   We have $\tau_i = 2\, \sqrt{w_i^2 - b^2}/v_T,$ where $b$ is the
distance of closest approach.
 This definition 
of the \ev\ \du\ is directly related to the time during
which the \ev\ can actually be monitored.
It also makes it clear that any innovations that improve the 
photometric sensitivity of the observing programs may be
important in increasing the observable \du\ of planet-lens
\ev s, thus increasing the chances that such \ev s will
be discovered and studied.
When we present the results of our simulations, we will
take $A_{min}=1.06.$

\subsection{Finite-Source-Size Effects}

Because the Einstein radius scales as the square root of the lens mass,
the Einstein radii of planet-mass lenses can be comparable to the projected
radius of the source star. With $D_S= 10$ kpc and $x= 0.9,$ $R_E=17.8 R_\odot$
($R_E=1.0 R_\odot$) for a Jupiter-mass (Earth-mass) lens. 
When the size of the source
becomes a significant fraction of the size of the Einstein ring, \fsse\ become
important, and the \mage\ is not well described by the formula of Eq.\ 1.
There are two important effects. 
First, the peak of the \lc\ becomes attenuated.
The larger the source size relative to the \er, the more
pronounced the attenuation.
The maximum magnification possible
for a disk of constant surface brightness, for example, is (Schneider,
Ehlers, \& Falco 1992)   
\begin{equation}
A_{max}={{\sqrt{R_S^2+4}}\over{R_S}}.
\end{equation}
This formula indicates that the
size of the source must be fairly large in order for the
peak \mage\ to be brought below the level of detectability,
particularly if the photometric
sensitivity is good. (Note that this applies to an isolated point mass;
a more exact
formula given by Gould \& Gaucherel 1997.)  
For example, a magnification of [$1.34$, $1.06$, $1.02$] can be achieved
even if the size of the source is $\sim [2.2, 5.7, 10.]  R_E$,
 Note that, 
when the lens is of Jupiter-mass (Earth-mass),
then a star with radius roughly equal to $180\ R_\odot$ ($10\, R_\odot$)
would fill a disk of radius $10\, R_E$. (We have used 
 $D_S=10$ kpc and $x=0.9.$)  
Thus,
in spite of the decrease in peak \mage, the large majority 
of stars in the source population can, when are lensed by \pl s in \wo s,
produce \ev s that should be detectable, at least with good photometry.

The second effect of finite source size on the \lc\ is to broaden
the width of the perturbed region of the light curve, since some portions
of the source may be significantly magnified well before the center of the
source
achieves its closest approach to the lens.
The \lc s can become almost flat-topped.
Thus, magnifications close to the peak value may be sustained during
the time it takes the source to travel several Einstein diameters,
and the event appears to
last longer than it would have had the lensed source been a point source.
This can be a boon to the detectability of \ev s due to
planet-mass lenses, since one of the greatest barriers to detection is
the short lifetime of the \ev .

\begin{figure}
\vspace {-1 true in}
\plotone{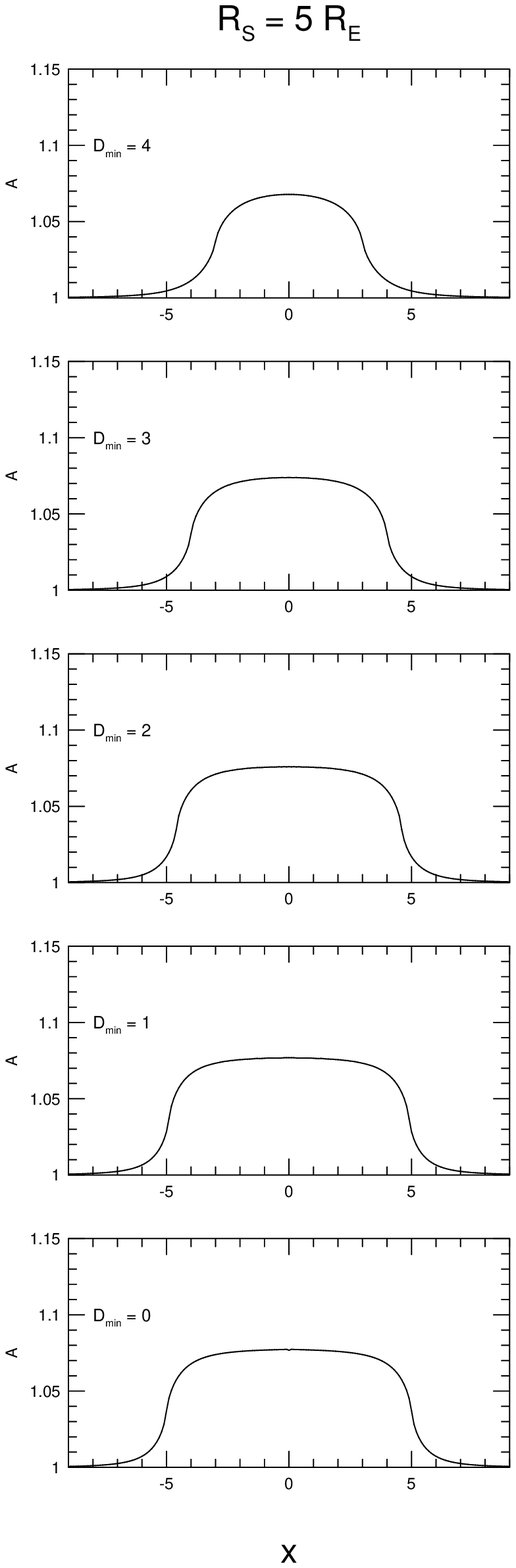}
\vspace {-.5 true in}
\caption{
In each panel, the projected size of the source in the lens plane is
$5\, R_{E,i}.$ The distance of closest approach, $D_{min}$, decreases from
$4\, R_{E,i}$ in the upper panel to the $0$ in the lower panel.
Note that the peak \mage\ is larger than $1.06$ in all cases.
Furthermore, if the measured
\ev\ \du\ is the time during which the \mage\ is larger than $1.06,$
then
these \ev s last up to   $2.5$
(bottom panel) times as long as they
would have, had the lensed source been a point source.
We have assumed that the disk of the lensed source has constant surface
brightness.
}
\end{figure}

\begin{figure}
\plotone{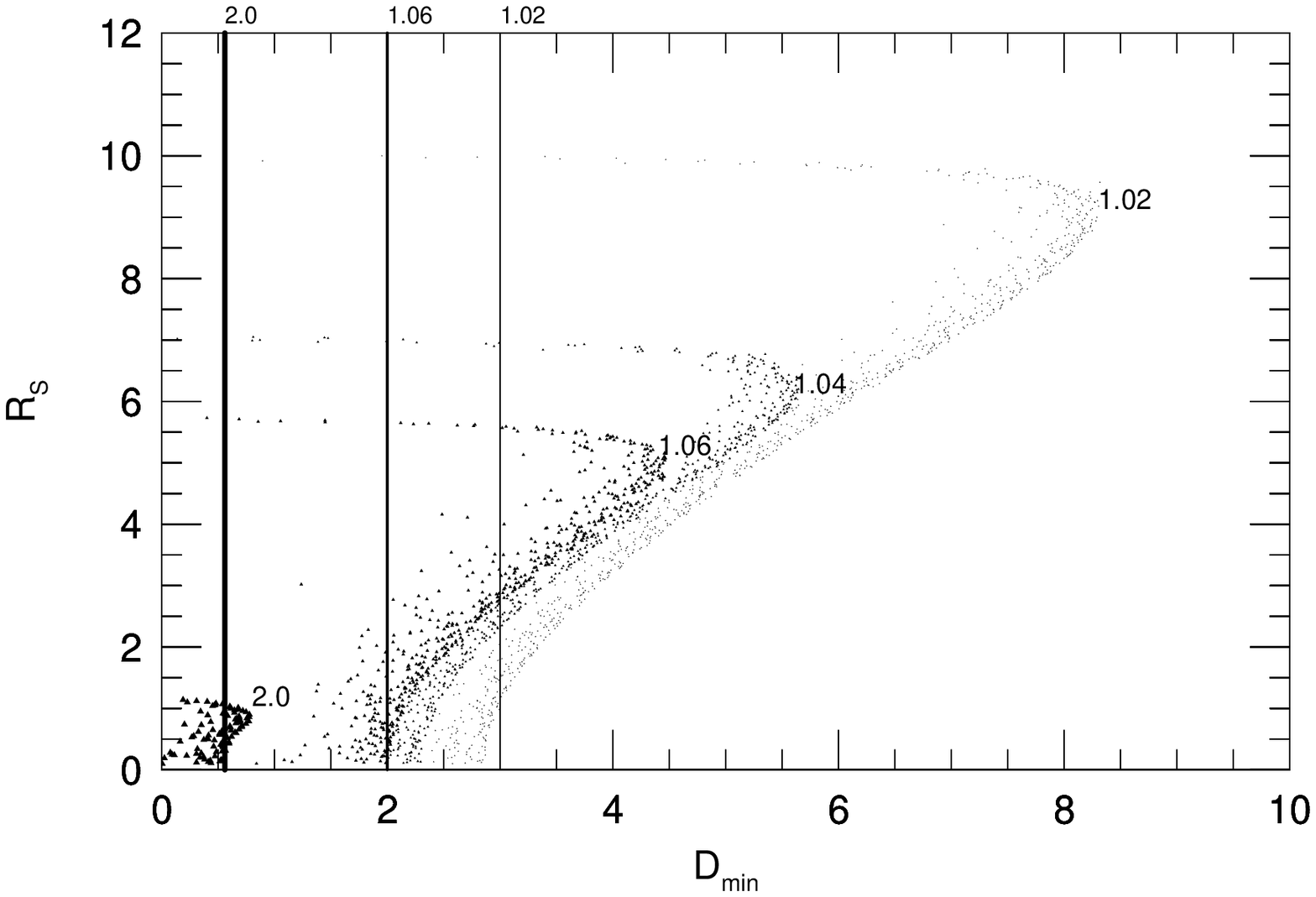}
\vspace {-4 true in}
\caption{
The source radius, $R_S,$ is
plotted against $D_{min},$ the distance of closest approach.
Both quantities are expressed in units of the Einstein ring radius.
Each point
corresponds to an event in which the peak \mage\ was roughly equal to
the value listed on the upper right of each curve.
For each value of $R_S,$ only the points with the largest values of
$D_{min}$ leading to the listed \mage\ are shown.
We have assumed that the disk of the lensed source has constant surface
brightness.
The value of $D_{min}$ for a point source is shown, for $A_{min}=
1.02, 1.06,$ and $2.0$ [vertical lines].
}
\end{figure}

Figure 1 displays a sequence of \lc s for the case $R_S = 5\, R_E.$
When comparing the different \lc s, note that the peak \mage\
does not change much, even as the distance of closest approach, $D_{min},$
changes from $0$ to $4\, R_E$. If the photometric sensitivity of the
observations was such that \ev s with peak \mage\ above $1.06$
could be detected, then all of these \ev s would be detected.
The primary difference as $D_{min}$ increases, is that the time \du\
of the observed \ev\ decreases. Even so, for $D_{min}$ as large as $4\, R_E,$
the
observed duration
is still roughly equal to the time a point-source \lc\ would remain above
$A= 1.06.$
The relationship between $D_{min}$ and $R_S$ is explored more systematically in
Figure 2.
The calculations used to produce these figures assumed that the
surface brightness of the lensed star was constant across the face of the
star. More realistic stellar profiles will change the details, but the
general features are robust. 
The implications
are that:
(1) depending on the distribution of source sizes,
the \dtn\ probabilities can be significantly increased by \fsse , and (2) \ev s
that can be detected can also generally be monitored for longer periods of
time.

Gould (1994) and Peng (1997) were able to find point-source fits
to \lc s associated with extended sources. This degeneracy of the
\lc\ shape is not expected to be problematic, however, when the source is
larger relative to the Einstein ring, as it will be in 
many of the cases relevant to the discovery of wide-orbit \pl s
(\rd\ 1998a). Moreover, even if there 
are \lc\  
degeneracies, these will not necessarily
prevent the identification of
individual \ev s as having been influenced by \fsse , since
spectral information can break such degeneracies. Real stellar
disks exhibit brightness profiles and spectra that have spatial
structure. When a star for which $r=R_S/R_{E,i}$ is not negligible is lensed,
the observed spectrum, and therefore the
\lc\ as observed at different wavelengths,
will be time-dependent. This effect 
can be used to study the lensed star.
(See, e.g., Witt \& Mao 1994,
Witt 1995, Gould 1994, Loeb \& Sasselov 1995,
Simmons, Willis \& Newsam 1995, Gould \& Welch 1996, 
Sasselov 1997, Heyrovsky \& Loeb 1997, and, for an observation,
Alcock {\it et al.} 1997.)
This time dependence also helps to confirm the interpretation of the
\ev\ as due to \ml\ of a source whose size has influenced the shape of the \lc .
Indeed, when the source is large relative to the Einsein ring,
it is as if a magnifying glass were scanning the face of the lensed star,
thus providing a good deal of information
about the source star and about the \ev . Spectral studies, even during
\shdn\ \ev s  can therefore be valuable
in breaking the degeneracy for individual \ev s,
as can multi-color photometry.  Even if spectral information is not
available, however, statistical analysis of the ensemble of
\ev s would be able to indicate that \fsse\ had played a role.
In particular, the relationship between peak magnitude and duration
would be different than it would be for a set of true Paczy\'nski \lc s;
for large sources, the peak magnification is closely related to
source size, while the duration continues to have a more direct
relationship to the
\dca .

\subsection{Resonant, Repeating, Overlap, and Isolated Events}
 
In the companion paper (\rd\ \& Scalzo 1998)
we provide an overview of how \pl-lens \ev s of different
types (resonant, wide \rpe s, wide isolated \ev s) 
complement each other.
 To orient the reader, we provide a thumbnail
sketch here. 

\noindent{\bf Orbital separation:\ } 
 When the separation, $a,$
 between the 
\cs\ and a \pl\ is much smaller 
 than the star's Einstein radius (typically $a < 0.5\ R_E$),
the large majority of \ev s in which the \ps\ serves as a lens
will be indistinguishable from stellar-lens \ev s in which an
isolated star serves as a lens. As $a$ increases, 
there is a marked increase in the fraction of stellar-lens \ev s 
displaying distinctive ``resonant" perturbations linked
to the presence of the \pl . 
As $a$ continues to increase, however,  
such \ev s again become rare: i.e., more distant \pl s are not in the \zres.

\noindent For \pl s located just beyond the \zres, 
the probability of \rpe s achieves
its maximum value.
In fact, if the photometric sensitivity is good, some \rpe s
(in which the source passes through the \lr s of both the star
and a \pl ), will exhibit
\lc s in which the \mage\ has not yet returned to baseline levels when the
influence of a second lens becomes apparent. We refer to this subset of
\rpe s as ``overlap" \ev s. 

\noindent Conceptually, overlap \ev s are most closely
related to repeating events, since they occur when the \lr s of two lenses are
transited by a distant source. Phenomenologically, however, they are ``isolated"
\ev s, in that the associated \lc s exhibit only a single excursion from
the baseline flux. If the photometry is sensitive
enough, there is no wait-time between \ev s, and this increases the
probability they will be identified, relative to the probability of detecting
other repeating \ev s.  Because they are relatively
long-lived, with their time duration keyed to the duration of
stellar-lens \ev s,
they are also more likely to be detected than
isolated \ev s of \shdn. 
In this paper we focus primarily on isolated events of \shdn , since
these will be most challenging to detect. Because they appear
as isolated \ev s, however, we do include
overlap events in the computations described and summarized in
\S 4.  

\noindent As  $a$ increases
further, all \rpe s display genuine repeats; i.e., they 
restart after a fall back to 
baseline following an initial event.
 The probability of \rpe s, which scales as $1/a,$ decreases, and
the probability of \is\ \ev s of \shdn\ increases. At large
separations, isolated \ev s of \shdn\ dominate.  

\noindent Thus, each type of \ev\ probes a different 
region of \ps s that act as lenses.   
(We note, however, that because different spatial orientations
are expected, the regions we study with each type of \ev\ can intersect
the regions explored by the others.)
Detections of each type of \ev\ therefore complement detections
of the others, with the ensemble of all \ev s providing
information about the spatial structures of and mass distributions
within \ps s. In addition, discoveries of
resonant and wide \ev s are subject to
different selection effects, allowing them to complement each other in another way.

\noindent{\bf The light curves:\ }
The \lc s associated with resonant \ev s are essentially stellar-lens
\lc s, exhibiting short-lived perturbations due to the presence of a \pl .
In principle, the mass ratio between the \pl\ and star, and the projected value of
their orbital separation, can be derived from the \lc ; in practice, there 
are degeneracies in the physical solutions that must be considered
(Gaudi 1997, Gaudi \& Gould 1997).
The \lc s associated with wide-orbit \ev s exhibit at least one peak that is of
relatively short duration, since at least one lens is a \pl\ and not a star. 
For \rpe s, as for resonant \ev s, 
the mass ratio and orbital separation can be 
constrained based on the \lc\ alone. As always, there are degeneracies,
but the fact that we see two separate \ec s in which one source is lensed
may help to resolve those degeneracies
 involving blending and finite source size.
For isolated \ev s, the \lc\ alone generally provides no
specific information about the orbital separation, and we
can only derive
a lower limit
on the distance between the central star and the \pl , expressed in terms of
the size of the star's Einstein radius. If, however, 
the spectral type of the \cs\
can be determined (which may be possible if it contributes a 
significant fraction of the baseline flux), and/or if \fsse\ are
important, then we can learn more even about planet lenses giving
rise to isolated \ev s. (See the discussion below.)  
In addition,
for some isolated \ev s,  the form of the \lc\ will deviate from the
standard form in a measurable
way.
In a very small fraction of cases
there may be caustic crossings, with the \lc\ form also significantly influenced
by \fsse . In others, the \lc\ will be well-fit by the formulae
of \rd\ \& Mao, allowing the projected orbital separation to be derived.

\noindent{\bf Novel aspects of planet detection via \ml :\ }     
In addition to the wide-orbit channel for detection, there are
other aspects of \pl\ detection via \ml\ that have been little-explored.
One of these has to do with follow-up work that can be done
for planet-lens \ev s, to learn more about the \ps\ that served as a lens.
While it has widely been assumed that little can be learned
about \ps s discovered through lensing, the combination of blending
and \fsse\ can play a powerful role in helping us to determine the
spectral type of the \cs\ and the masses of any \pl s involved in the lensing
\ev . We touch on this here (and in \rd\ \& Scalzo 1997, 1998),
but discuss the possibilities more fully in \rd\ 1998b.
The relevance to this paper is that it is possible that 
observations designed to complement \lc\ studies can help us
to learn more about isolated \ev s of \shdn. The additional information
can, under favorable circumstances, include the spectral type of the \cs\ and the
mass of the \pl\ lens. 
Another
interesting question is whether  the \pl s discovered via \ml\
are likely to harbor life; this is discussed in \rd\ (1998a).  
Finally, we note that it may be possible to detect 
fine structure within the lensing planetary system, in the form of moons 
orbiting \pl s and asteroid or cometary belts (\rd\ \& Keeton 1998).
\footnote{All of these topics, together with both isolated and repeating wide-orbit
\ev s, are covered in \rd\ \& Scalzo 1997 as well. The referee of that
paper suggested that it would be preferable to discuss the separate ideas
in separate  
papers, thus leading to the sequence
of  papers described here. The collection of individual
papers includes some new work.}

\section{Do We Expect To Find Planets in Wide Orbits?}

The excitement about resonant lensing by planets was fueled, at least in part, by a
wonderful coincidence. Gould \& Loeb (1992) noted  that,
if a system identical to our Solar System happened
to be located halfway between our position and the center of the Bulge, and if
the system were viewed face-on, the separation of the planet corresponding to
Jupiter from the system's star would be very close to the value of $R_E$
associated with the mass of the star.  That is, Jupiter would be in the 
\zres . There are,
however, two reasons to be cautious about using this example to limit the
search for planets to those in resonant orbits.  First, we do not know that
the spatial relationship between the Sun and Jupiter is an example of a
universal property of planetary systems. 
Mindful of this, Bennett \& Rhie (1996) have
constructed a ``power-of-2'' planetary model, in which the separation between
each planet and the central star increases by a factor of two for each
successive planet.  In such a model, most planetary systems can be expected to
contain one planet in a resonant orbit.
Each such \ps\ contains several (${\cal O}(10)$) \pl s in \wo s. 
A second reason to avoid limiting the microlensing searches to resonant
planets is that the value of the Einstein radius depends not only on the
stellar mass, but also on the relative positions of both source and lens to
the observer.

In Figure 3 we consider planets with orbital separation
$a=2\, a_w = 3\, R_E$. For these systems we show
the relationship between the stellar mass, $M$, and $x$, when the
orbital period is fixed.
We have
assumed that the orbital plane is the same as the lens plane,
and show those values of $x$ for which both lens
and lensed source are located
in the source galaxy (the Bulge, the Magellanic Clouds, or M31). If the
stellar mass is in the range from $0.1$ to $10\, M_\odot,$ the orbital
periods of \pl s in \wo s range from a few years to a few hundred
years, with larger values more typical for more distant galaxies.

\begin{figure}
\vspace{-1 true in}
\plotone{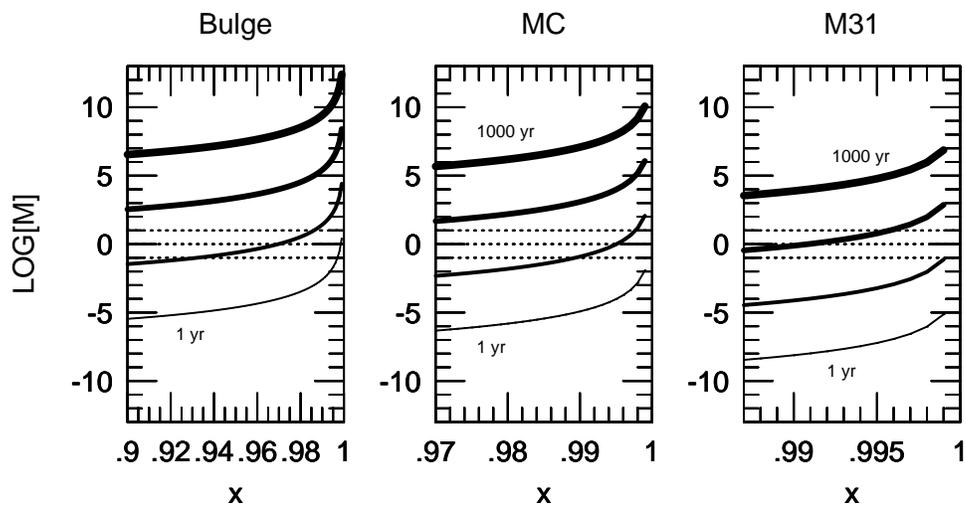}
\vspace{-4 true in}
\caption{
Each curve corresponds to a fixed value
of the orbital period; shown are
curves for 1 yr, 10 yrs, 100 yrs, 1000 yrs.
Each curve illustrates the relationship between the stellar mass, $M,$ 
and $x$ for a 
\pl\ in a \wo, with $a=3\, R_E$. In each panel the range of the
variable $x$ corresponds to lenses located within the 
same galaxy (Bulge, Magellanic Clouds, M31 [left to right])
as the lensed stars. 
Note that the range of masses and orbital periods best explored via
wide lensing events is different for different source galaxies.
}

\end{figure}

\tightenlines
\begin{deluxetable}{llrlll}
\label{visibility}
\scriptsize
\tablecaption{Planets in wide and resonant orbits in known and model systems;
\hfil\break $D_s = 10$ kpc, $x = 0.9$.}
\tablehead{\colhead{Planetary system} & {Planet} &
   \colhead{$a$ \tablenotemark{(1)} } &
   \colhead{$\%_{close}$ \tablenotemark{(2)} } &
   \colhead{$\%_{res}$ \tablenotemark{(3)} } &
   \colhead{$\%_{wide}$ \tablenotemark{(4)} }} 
\startdata
{\bf Solar system:} 
& Jupiter & 5.2 & \, 5.9  & 21.0 & 73.1 \\
& Saturn  & 9.5 & \, 1.7  & \, 4.6  & 93.7 \\
& Uranus  & 19.2 &\, 0.4  & \, 1.0  & 98.6 \\
& Neptune & 30.1 &\, 0.2  & \, 0.4  & 99.4 \\
& Pluto   & 39.8 &\, 0.1  & \, 0.2  & 99.7 \\
\hline
{\bf Known extrasolar systems:} 
& 55 Cnc       & 4.0 & 11.0 & 89.0 & \,\, 0.0 \\
& HD 29587     & 2.5 & 37.0 & 63.0 & \,\, 0.0 \\
& PSR B1620-26 & 38 & \,\, 0.1 & \,\, 0.3 & 99.6 \\
& Gl 229b &    40 & \,\, 0.1 & \,\, 0.2 &  99.7 \\
\hline
{\bf Theoretical power-of-2 system:  \tablenotemark{(5)}}
& 1 & 4.8 & \,\, 7.0 & 27.0 & 66.0 \\
& 2 & 9.6 & \,\, 1.6 & \,\, 4.5 & 93.9 \\
& 3 & 19.2 & \,\, 0.4 & \,\, 1.0 & 98.6 \\
\hline
{\bf Theoretical power-of-3 system:  \tablenotemark{(6)}}
& 1 & 4.8 & \,\, 7.0 & 27.0 & 66.0 \\
& 2 & 14.4 & \,\, 0.7 & \,\, 1.9 & 97.4 \\
& 3 & 43.2 & \,\, 0.1 & \,\, 0.2 & 99.7 \\
\hline
\enddata
\tablenotetext{(1)}{Separation of the planet from the central star
in astronomical units (AU). We
consider circular orbits. 
See the text for the criteria used to select the \pl s listed here.} 
\tablenotetext{(2)}{Percentage of the time during which the planet is
``close'' ($a < 0.8 R_E$), averaged over all inclinations.}
\tablenotetext{(3)}{Percentage of the time during which the planet is
``resonant'' ($ 0.8 R_E < a < 1.5 R_E$), averaged over all inclinations.}
\tablenotetext{(4)}{Percentage of the time during which the planet is
``wide'' ($a > 1.5 R_E$), averaged over all inclinations.
In the power-of-2 and power-of-3 models,
all planets beyond the third had a probability greater than $99\%$
of being located more than $1.5\, R_E$ from the central star.}
 \tablenotetext{(5)}{The orbital separation increases
 by a factor of $2$ for each successive \pl .
When we averaged over possible positions for the innermost \pl\
 (from 0.3 AU to 2.0 AU), keeping the number of planets constant at 12,
 we found that
 1.90 planets on average were close; 0.925 planets on average were
 resonant; and 9.17 planets on average were wide.}
 \tablenotetext{(6)}{
The orbital separation increases
 by a factor of $3$ for each successive \pl .
 When we averaged over possible positions for the innermost \pl\ 
  (from 0.3 AU to 2.0 AU), keeping the number of planets constant at 7,
we found that 
1.41 planets on average were close; 0.535 planets on average were
 resonant; and 5.06 planets on average were wide.}
\end{deluxetable}

In addition, the projected value of the separation depends on the
angle, $\alpha$, between the normal to the orbital plane and our line of
sight.
Consider an orbit with $a > a_w$. Let $\alpha$ represent the angle between
the normal to the plane of the orbit and the normal to the lens plane.
For $\alpha < \alpha_{min}$, where
\begin{equation}
\alpha_{min} = \cos^{-1}(a_w/a),
\end{equation}
the projected value of the orbital separation is always greater than $a_w$.
If $\alpha > \alpha_{min}$, then for a circular orbit, the fraction of time
for which the projected separation is greater than $a_w$ (so that the planet
can be viewed as wide) is
\begin{equation}
f = \frac{2}{\pi}
    \cos^{-1} \left(
            \frac{\sqrt{\cos^2 \alpha_{min} - \cos^2 \alpha}}{\sin \alpha}
            \right).  \label{frac-crw}
\end{equation}

\noindent
If the orbit is eccentric, then the separation between the planet and star will
be wide for an even greater fraction of the time.

Thus, even if planetary systems had uniform properties, the fact that
they are located at different spatial positions and are tilted at different
angles relative to our line of sight would still not favor resonant orbits over
others.  Bennett \& Rhie's power-of-2 model, or something like it, is
necessary to ensure that most planetary systems have one planet in the zone
($a_c < a < a_w$) associated with resonant lensing.  This is because
$a_w/a_c \approx 2$; thus, if the $i^{th}$ planet has separation $a_i$ from
the central star, and the power-of-2 system is inclined so that
$a_i \cos \alpha \approx a_c$, we have $a_{i+1} \cos \alpha \approx a_w.$

The known planetary and brown-dwarf binary systems are considered in Table 1,
together with a ``power of 2'' model and a ``power of 3'' model.
The known systems included in the table were selected from the
Encyclopedia of Extrasolar Planets 
(accessible from http://www.obspm.fr/planets)
Two
criteria were used: (1) if the \ps\ were placed in the Bulge and viewed face-on,
the planet (or brown dwarf) listed had to be in an orbit that
would either be wide or located in the \zres , (2) the existence of the \pl\
or brown dwarf 
needed to be listed as ``confirmed".
 In the power-of-2 and power-of-3
model, the first planet listed is the innermost \pl\ that 
would be in a \wo, were the \ps\  to be placed in the Bulge and viewed face-on.
We note that the next 
\pl\ inward would have a good chance of being viewed in the \zres .
Indeed for both of these theoretical models, there is some
ambiguity associated with the arbitrarily-chosen position of the
closest \pl\ to the central star. Because of this, we carried out
a set of calculations in which we averaged over the position of the innermost
\pl .
We found that, if we truncate the
radius of the \ps\ at $\sim 10^{17}$ cm, 
there are on average $\sim 9-10$ \pl s in \wo s for
every \pl\ located in the \zres.    
To derive the numbers shown in the table, we averaged over inclination angle
$\alpha$, and used equations (6) and (7) 
to determine the fraction of time any given planet would be in a wide orbit.
As above,
we have assumed that the orbits are circular.

The considerations in this section show that, under a range of
reasonable assumptions, planets in wide orbits 
are very likely to exist. 
In fact the number of planets in wide orbits may be as much as
an order of magnitude larger than the number of planets in resonant
orbits. The relative contribution of these
two classes of planet lenses to the detection rate depends on
our ability to identify the associated events.

\section{The Rate of Detectable Isolated Planet-Lens Events}

\subsection{Normalization of Event Rates}  

We will be interested in the rate at which detectable \ev s occur,
 and how that rate is influenced by changes in the detection strategy.
The Einstein radius of the 
central star provides a convenient normalization.
The rate of \ec s in which the central star serves as a lens,
with the magnification due to the central star achieving a peak value 
of at least $1.34$ ($A_{min} = 1.34$), is proportional to 
$2\, R_E.$ Because the width of the central star's \lr\  (when $A_{min} = 1.34$) 
will play an important role, we define $w_{0,0}$ to be equal to $R_E$.  
If the photometric sensitivity is such that $A_{min}$ can be smaller than
$1.34$ (i.e., $w_i > R_E,$ $n>1$), so that the effective width of \ev s involving the central
star is larger ($w_0 > w_{0,0}$), the rate of all detected \ev s,
including those due to the central star,
will increase. 
Nevertheless, 
 $2\, w_{0,0}\, $ is a convenient normalization constant,
because it allows us to compare the rates we compute to
 the presently-measured rate of \ev s.
This is as follows.
Along the direction to the
Bulge, \ev s 
 are being discovered by the MACHO team at a rate of
roughly $50$ per year.  
The present detection criteria used by the MACHO team
are strict: $A_{min}=1.58$, corresponding to $n$ (in Eq.\ 3)
being set equal to $\sim 0.76.$   In addition, other teams,
surveying different fields with somewhat different strategies,
are also surveying the Bulge. For example, the OGLE team,
which in its first incarnation discovered $\sim 12$ Bulge \ev s,
has recently
brought a new telescope on-line (Udalski, Kubiak \& Szymanski 
1997; Paczy\'nski {\it et al.} 1997). 
It is therefore
reasonable to assume $75-100$ \ev s of the type
we use for our normalization per year along the direction to the 
Bulge. Thus, because we use $2\, w_{0,0}$ to normalize our results, when
we find  
that a particular detection
strategy leads to a rate of detectable 
\ev s of a certain type (e.g., \ev s with $2$ repetitions)
equal to $p\, \%,$ this means that between $0.75\, p$ and $p$ such \ev s
could be discovered per year along the direction to the Bulge.

\subsection{The Rate of Isolated Events}

When a wide planetary system serves as a lens,
the most common type of event
is one in which the track of the source passes through the Einstein ring of
a single mass.  Because the rate of \ec s associated with a given lens scales
as the square root of its mass, \ec s in which the star serves as a lens are
the most frequent. Let $P_1$ represent 
the rate of \is\ \shdn\ \ev s due to \pl\ lenses. 
$P_1$ is proportional to the linear dimensions of the \lr .   
\begin{equation}
P_1 \sim 2\, \sum_{i=1}^N w_i \sim 2\, \sum_{i=1}^N n_i\, R_{E,i} \sim
\sum_{i=1}^N n_i \sqrt{q_i}, 
\end{equation}
where the sum is over all of the system's \pl s, and $q_i = m_i/m_\ast.$ 
When \fsse\ are not important, $w_i = n\, R_{E,i}.$ 
The normalization described in \S 4.1, yields a normalized rate of isolated
\pl-lens \ev s:
\begin{equation}
P_1 = \sum_{i=1}^N n_i \sqrt{q_i} = n\, \sum_{i=1}^{N}  \sqrt{q_i},
\end{equation}
Note that, because the \ev\ rate scales with $n,$ improvements in
photometric sensitivity play an important role in optimizing the
rate of detectable \ev s, even when \fsse\ are not important. 
When \fsse\ do play a role, the increase in rate can be even
larger for some of the terms on the right hand side of Equation 8. 

The expression for $P_1$ given in Eq.\ 9 
overcounts \is\ \ev s slightly, because the
rate of \rpe s  
must be subtracted from it.  
Because, however, isolated \ev s involving a specific lens are generally 
$1-2$ orders of
magnitude more common than repeating \ev s involving the same lens, 
(\rd\ \& Mao 1996; \rd\ \& Scalzo 1997, 1998), $P_1$ generally provides
a good estimate of the rate of isolated events involving \pl s.

The time duration of isolated
planetary-lens events, relative to the time duration of stellar-lens events,
also scales as $\sqrt{q_i}$.
The implication is that 
a distribution of events due to lensing
by stars with wide-orbit planets
is necessarily accompanied by a distribution
of shorter-duration events.  The fraction of events in the latter
distribution is proportional to the average value of $\sqrt{q}$, and the position of
the peak or peaks also provides a measure of the mass ratios typical of
planetary systems.

Possible forms of the distribution of \ev\ \du s are  
illustrated in Figures 4 and 5. (The rates corresponding to the
integrated area under the curves for \pl-lens \ev s are discussed
in \S 4.3.)  These figures show results
from a set of Monte-Carlo simulations in which 3 types of  \ps s 
were placed in the Bulge and served as  lenses for 
more distant Bulge stars ($D_S=10$ kpc, $x=0.9$). The details of the
simulations and a more complete description of the results are provided
in the companion paper (\rd\ \& Scalzo 1998). 
Shown are the results derived when the \ps\ lens
was (a) identical to the Solar System (top panel),
(b) a power-of-$3$ system (middle panel), and 
(c) a power-of-$2$ system (bottom panel).
(Note that, in the power-of-$n$ models, the orbital separation
between \pl s and the \cs\ increases by a factor of $n$
proceeding outward from the \cs .) 
Each \ps\ was considered as a lens in numerical experiments that
generated thousands of random orientations of the orbital plane
with respect to the lens plane,
and for each, considered the passage of source tracks coming in at
random angles, with velocities chosen from a Gaussian distribution
centered at $150$ km/s, with width equal to $50$ km/s. 

We show the (unnormalized) distribution of \ev\ \du s, defining the \du\ to be
the time during which the \mage\ was above $A_{min}=1.06$. 
In all cases, a long-duration peak in the distribution, due to
lensing by the central star, is necessarily accompanied by
one or more smaller peaks due to \shdn\ planet-lens \ev s. 
The difference between Figure 4 and Figure 5 
is that different sets of criteria were 
used to determine whether an \ev\ is detectable.  
Criteria A (Figure 4)
require the \lc\ to exhibit magnification greater than
	$A = 1.34$ (source-lens separation less than $1.0 R_E$) for at least
		1 day in order for the lensing \ev\ to be detected.
Criteria C (Figure 5) require only that
the magnification be greater than $1.06$ (source-lens separation 
less than 
	$2.0 R_E$), 
 and no minimum event duration is required.
Criteria C are more ambitious, and allow the effects of lensing
by several
\pl s to be seen. In the case of the Solar System, there are
peaks due to lensing by each of the outer \pl s, although some
of the structure is blurred as the peaks run together.
It is interesting to note that most of the peak structure actually
comes from Saturn and the planets beyond it, rather than from Jupiter.
The reason for this is that the separation between Jupiter and the 
sun is so close to $a_w$ that, for some orientations of the 
orbital plane with respect to the lens plane, Jupiter
was in the \zres\ at least part of the time. Furthermore, even when it
was in a wide orbit, its \lr\ sometimes overlapped that of the 
Sun. That is, before the \mage\ could slip below $1.06$ after
Jupiter served as a lens, the track of the source would already
be in the \lr\ of the Sun (or {\it vice versa}). The result would
be one long \ev\ with the \lc\ exhibiting two peaks. 
The dotted lines correspond to such ``overlap" events. 
Although the source track passes through the \lr s of two lenses,
there is just one continuous deviation from baseline. Overlap
\ev s appear to be isolated \ev s with unusual morphologies that
last  for times comparable to the durations of stellar-lens \ev s.  
Overlap \ev s are discussed in the companion paper 
(\rd\ \& Scalzo 1998). Like isolated \ev s of \shdn,
overlap \ev s are necessary adjuncts of 
the presence of \w-o\ 
\pl s. 
They are a feature of the power-of-$n$ simulations as well,
when there was often an overlap between the \lr\ of the \cs\ and the innermost 
wide \pl . 

In the power-of-$2$ and power-of-$3$ models,
all of the \pl s were assumed to be of Jupiter mass, so there
is just one statistically significant peak due to \pl\ lenses;
the area under this peak is large, however, because it
includes the contributions from all of the systems' wide \pl s. 
\begin{figure}
\vspace{-1 true in}
\plotone{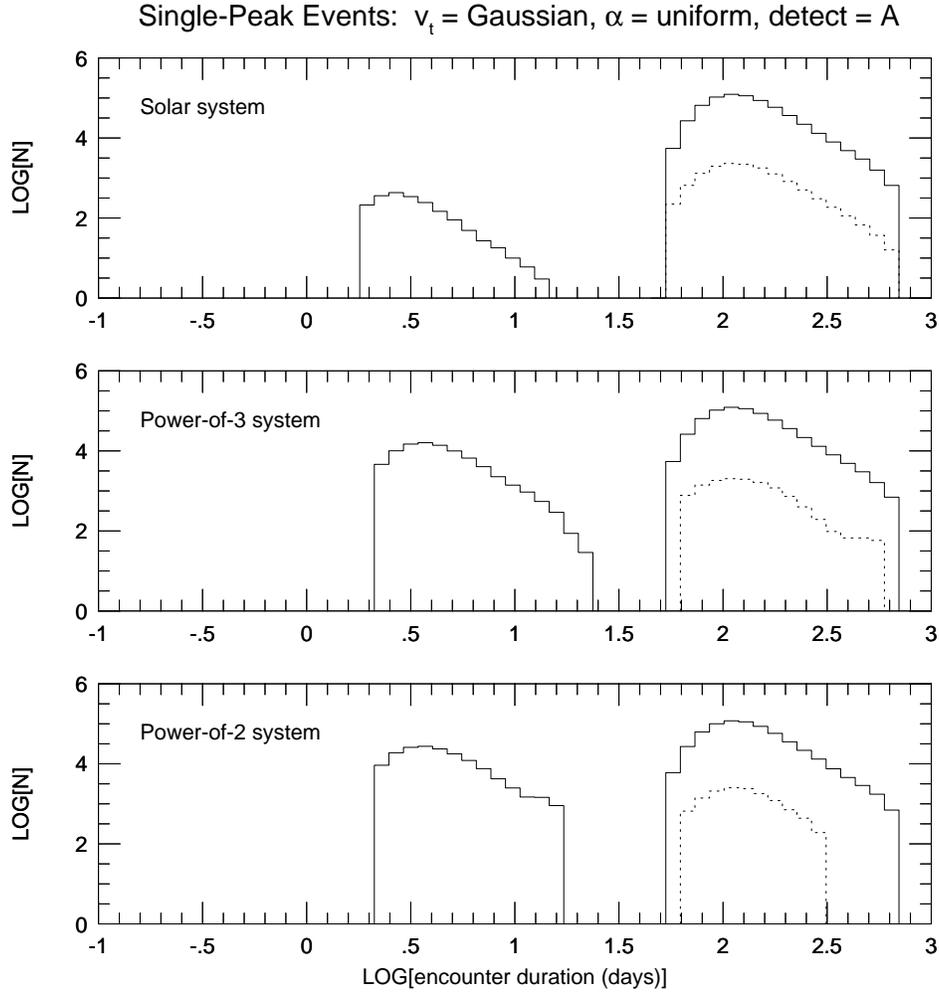}
\vspace{-2 true in}
\caption{Distributions of durations for isolated (non-repeating) events
using the detection criteria of set A (the \lc\ must
exhibit magnification greater than
	$A = 1.34$ [source-lens separation less than $1.0 R_E$] for at least
			1 day in order for the lensing \ev\ to be detected).
In each panel, the peak on the left is due
to isolated 
planet-lens events, and the peak on the right is due to
the central star.  Overlap events, produced almost exclusively by
\ec s involving the innermost wide planet and the central star,
are shown by the dotted lines. (See \rd\ \& Scalzo 1998.) 
The truncation of the large-duration end of the overlap distributions
in the power-of-3 and power-of-2 models is due solely
to the poorer statistics achieved in the simulations of these models;
all of the distributions were scaled to the linear sampling achieved
for the Solar System. 
In the Solar System (power-of-$3$, 
power-of-$2$) model (top panel [middle panel, bottom panel]),
there were $0.3$ ($13.0$, $23.7$) isolated \ev s of \shdn\ per year and 
$2.1$  ($1.9$,    
$2.2$) overlap
\ev s. }

\end{figure}
 
 \begin{figure}
 \vspace{-1 true in}
 \plotone{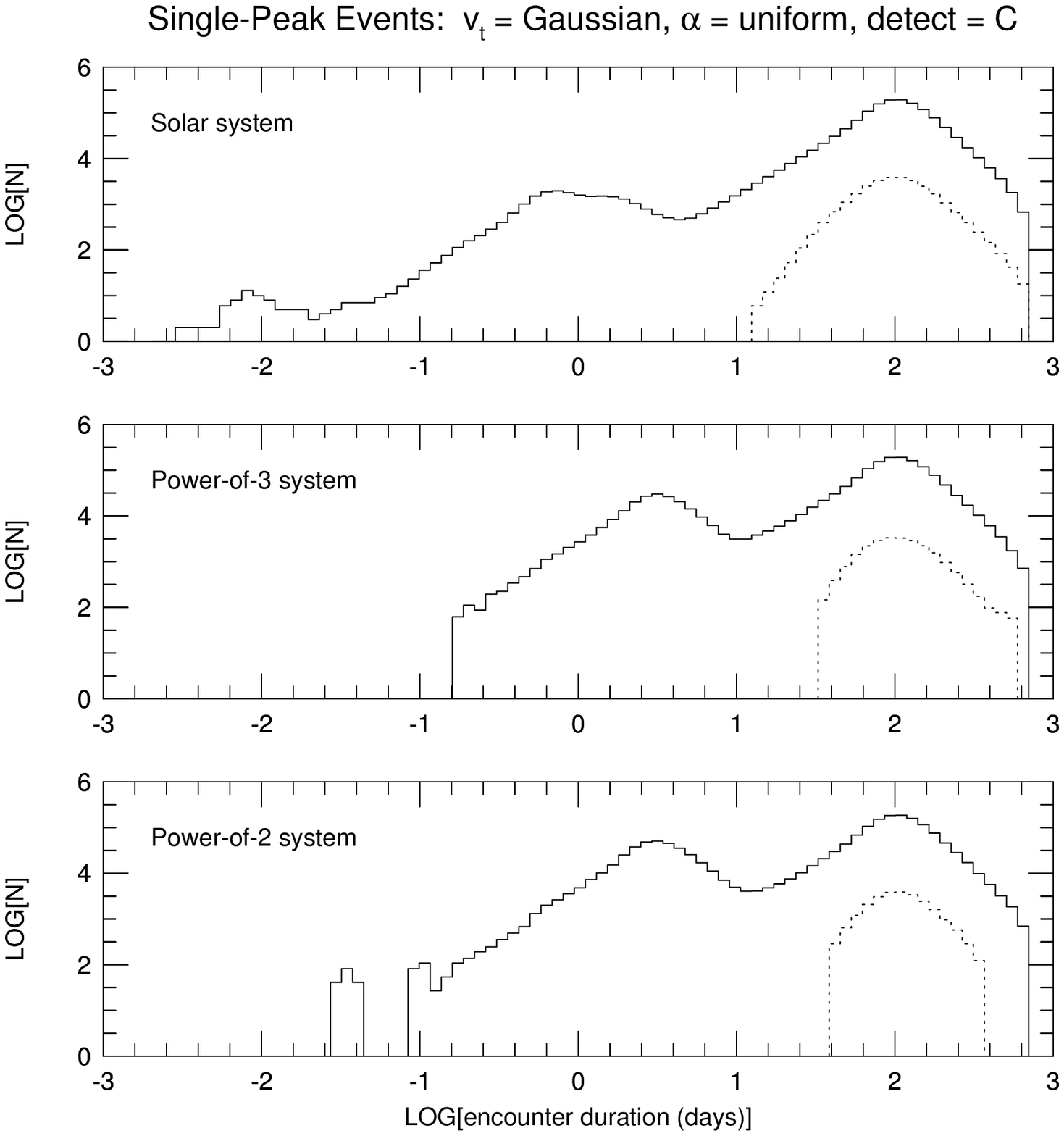}
 \vspace{-2 true in}
 \caption{Distributions of durations for isolated (non-repeating) events
 using the detection criteria of set C
 (the source-lens separation must be smaller than 
	 $2.0 R_E$, and no minimum event duration is required).
	    Here,
 because an \ec\ did not have to have a minimum \du\ in order to be detected,
  the peaks tend to
  have low-duration
  tails which can run together.
  The right-most peak in each panel is due to the central star of the system,
  with overlap events involving the innermost wide planet being shown by
  dotted lines (\rd\ \& Scalzo 1998); 
  the peaks on the left are due to planetary encounters.
  The central feature in the top panel is a superposition of peaks corresponding
  to Uranus, Neptune, and Saturn; the encounters due to Saturn are visible as a
  shoulder at about 1.6 days. 
In the Solar System (power-of-$3$,
power-of-$2$) model (top panel [middle panel, bottom panel]),
there were $3.1$ ($30.4$, $55.9$) isolated \ev s of \shdn\ per year and 
$4.2$  ($3.6$,   
$4.0$) overlap
\ev s. }
  \end{figure}

\subsection{Rates of Events Associated with Known Systems}

The \ps\ we know best is our own. Not only do we know that the Sun is accompanied
by a \pl , but we know that there is a set of \pl s distributed over several
tens of astronomical units (au). 
If a \ps\ identical to the Solar System is located in the Bulge, the Magellanic
Clouds, or in M31, several of its \pl s (generally Jupiter and all of the outer \pl s)
are in \wo s most of the time, for most orientations
of the ecliptic on the sky. If a lensing \ev\ with a solar mass lens lasts for 100
days, then an event in which [Jupiter, Saturn, Uranus, Neptune, Pluto] serves as
a lens lasts for [$3.08$, $1.68$, $0.66$, $0.71$, $0.01$] days.
If all stars were as massive as the sun, and were accompanied by a \ps\
identical to ours, then the durations given in the previous sentence 
could be translated directly into the percentage of \ev s in which each of the \pl s
in turn could be expected to serve as a lens. The number of \pl-lens \ev s per year
would be roughly $2-4$ in Baade's window (for $A_{min}=1.34,$) and $\sim 5-10$ 
(for $A_{min}=1.06).$ 
With the present
\mo\ strategy, \ev s lasting $1$ day or less would not be subjected to
detailed enough \mo\ to unambiguously identify them as \ml\ \ev s. 

To make contact with specific detection criteria, we refer to the 
Monte-Carlo calculations corresponding to Figures 4 and 5. These yield a rate
of isolated \ev s of only $0.3$ per year for the Solar System, when
detection criteria A are used. This is because The \lr\ of Jupiter
actually overlaps that of the Sun for most 
orientations of the orbital plane on the sky.  Thus, Jupiter is
detected through overlap \ev s, which occur at a rate of $2.2$ per year 
when detection criteria A are employed. The only other outer \pl\ that
can produce an \ev\ lasting longer than a day is Saturn, and not every
\ev\ due to Saturn lasts this long. Most of the overlap \ev s should be
detectable, and should also provide some evidence of the presence of Jupiter.
The set of detection criteria C leads to $3.1$ \shdn\ isolated \ev s
per year and $4.2$ overlap \ev s. 
Thus, over a period of $10$ years, dozens of \pl-lens
\ev s could be detected.

The \ev\ \du s and rates scale with $w_i$, and $w_i$ depends on \fsse . We
therefore note that, if the \re\ of an Earth-mass \pl\ is $\sim 1.0\, R_\odot,$  
then the \re\ of [Jupiter, Saturn, Uranus, Neptune, Pluto] is 
[$17.8$, $9.8$, $4.1$, $3.8$, $0.04$] $R_\odot$. Pluto would very likely not be
discovered via \ml . Jupiter, Saturn, Uranus, Neptune 
would be detectable. Furthermore, were the lensed star
to be massive and/or evolving, the time duration of the
\ev s could be significantly extended by \fsse .
Thus, depending on the distribution of radii among the
source population, the rate of detectable \ev s could be significantly
higher than given above. 

We now know of \ps s in addition to our own Solar System.
Among other systems in which the presence of a brown dwarf or planet-mass
is confirmed, there are two in which the brown dwarf or planet is in an orbit
that would be wide, were the system to be located in the Bulge and serve as a lens
for a more distant Bulge star ($D_S = 10$ kpc, $x= 0.9$).
These are Gl 229 (a brown dwarf system) and PSR B1620-26. If all stars 
had similar companions to the brown dwarf in GL 229 
(the \pl\ in PSR B1620-26), then      
there would be $\sim 20$ ($10$) \is\ \shdn\ \ev s per year along the
direction to the Bulge (for $A_{min}=1.34,$) 
and $\sim 40$ ($20$) \is\ \shdn\ \ev s per year (for $A_{min}=1.06).$  

It is simply not known whether our \ps\ or any of the others which we have
now begun to study are typical of \ps s elsewhere in our Galaxy or in other galaxies.
It may therefore be instructive to consider theoretical 
constructs. In the power-of-$n$ model, we assume that there is a \pl\
of mass $m$ located within $1$ au ($a_0 < 1$ au) of the \cs , and other \pl s located  
at orbits of radii  $n^i a_0$ for $i$ ranging from $1$ to the maximum value
consistent with a bound orbit. For the power-of-$2, 3, 4$ model,
with $A_{min} = 1.34\, (1.06),$ there would be 
$\sim 26, 15, 13 \, \sqrt{m/m_J}$ ($52, 30, 26 \, \sqrt{m/m_J}$) 
\is\ \shdn\ \ev s per year toward the Bulge per year.
The results of Monte Carlo simulations are consistent with these
predictions; for the power-of-$2$ and
power-of-$3$ models the rates are given in the captions to Figures
4 and 5.
    
The common factor in all these results is that the \ev\ rates
are large enough that, if \ps s are a common phenomenon,
we should be able to see evidence of \ps s
in the distributions of \ev\ \du s.

\subsection{The Detection and Identification of Isolated Planet-Lens Events}

Are the observing teams capable of discovering
\ev s of such short durations that the lens could be a \pl ?
If, so, and if 
planet-mass objects do serve as lenses, 
will we be able to identify
the associated peak of \shdn\ \ev s in the distribution 
of \ev\ durations? 
This would allow us to minimize contamination of the \pl-lens
signal due to the short-duration tail
associated with stellar lenses.
Finally, and perhaps most interesting,
can some individual  isolated events be reliably identified as \pl-lens \ev s?
If so, then we can eliminate some of the possible contamination due to
lensing by low-mass MACHOs.

\subsubsection{Observations of Short-Duration Events}

The shortest
\ev\ on record lasted approximately $2$ days. 
To better understand what is known so far, we have considered \ev s along the
direction to the Bulge, since the MACHO team has already accumulated a store of
over $150$ Bulge \ev s about which some information is publicly available.
Specifically, we have used their paper on $45$ Bulge \ev s monitored in 
1993 (Alcock {\it et al.} 1997a),
and their ``alert" web pages (http://darkstar.astro.washington.edu),
which list approximate \du s for many of the 136 Bulge \ev s observed
during 1995, 1996, and 1997. Altogether we found (up to October 1997)
 $148$ \ev s  
(a) which  are apparently 
due to lensing by a point-mass, and  (b) 
for which approximate \du s are available. (We note however, that
the \du s posted on the web site come from fits that may be refined in the future.)   
The \dbn\ of \ev\ \du s is plotted in Figure 6. Note the appearance of
the \shdn\ peak. 

\begin{figure}
\plotone{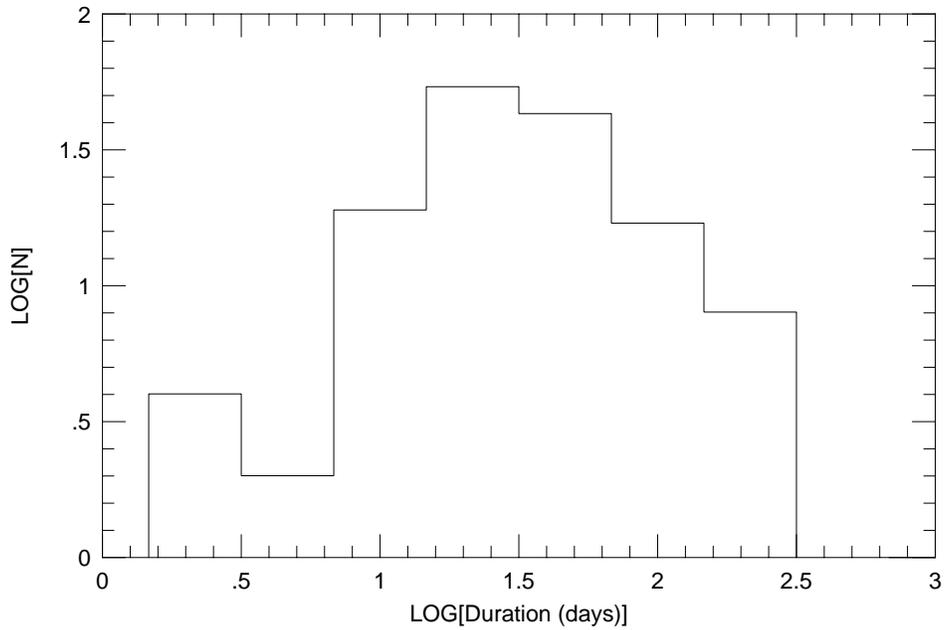}
\vspace {-4 true in}
\caption{
Distribution of \ev\ \du s for the MACHO team's set of Bulge \ev s.
Included is data from 1993 (Alcock {\it et al} 1997a), and the ``alert"
web page data for \ev s which occurred in 1995, 1996, and 1997.
All 148 \ev s which were not listed as binary \ev s, and for which
estimates of \ev\ \du s had been, made are included. Note that the
definition of duration used here is $(2\, R_E/v_T)$. 
}
\end{figure}

In light of Figure 6 we may ask whether the MACHO team has already
begun to discover concrete evidence for a signature due to \pl s in the
Bulge. This may be the case. (In fact 
Bennett {\it et al.} [1996a] have conjectured that one short \ev\ may indeed
be due to a \pl\ in a \wo .) 
We emphasize, however, that a good deal of
further work would be required to clearly establish it (\S 4.6, \rd\ \&
Scalzo 1997, 1998, \rd\ 1998a,b). 
One certain conclusion we can draw from the MACHO Bulge data is that
the team has proved that it has the ability to detect \shdn\ \ev s.  
This is encouraging, since 
the detection strategies presently used are not optimized for the
discovery of 
short time-scale \ev s.

\subsubsection{Identifying Peaks In The Distribution of Short Events}

Consider the distribution of \ev\ \du s. 
Is it likely that a peak due  to the presence of lenses with small masses
will be hidden by or confused with
 the short-duration tail of \ec s due to lensing by
stars?

Three circumstances can make a stellar-lens \ec\ have a short
measured \du .

(1) The relative velocity $v_T$ an observer would measure between the
source and the lens may be exceptionally large.  Since, however, the
distribution of transverse velocities for planet-lens and stellar-lens
\ec s should be roughly the same,
we should be able to disentangle velocity effects, at least
for statistical samples of \ec s.

(2)
The track of the source may just graze the lens' Einstein ring.  In this
case, the peak value of the magnification allows us to measure
the distance of closest approach,  $D_{min}$, 
and to determine
that, although short, the observed \ec\ was nevertheless
due to the presence of a lens with a larger Einstein ring
than expected for a planet.
These corrections are implemented by the observing teams.

(3) Light from the lensed source could be blended with light
from other sources along the line of sight. If,
however, the effects of blending are dramatic enough to shorten the
\du\ of an \ev\ significantly, then we should
be able to detect evidence for the blending in other ways, and to
subtract its effects from the event before comparing its \du\ with
that of other \ev s.
Because blending can provide a valuable tool for the study of planet-lenses,
we discuss it in more detail in separate papers (\rd 1998a,b).

We conclude that, if the present generation of
\mo\ \pr s discover a set of isolated \pl-lens \ev s,
they should be able  
to reliably 
separate the signal from the contribution due to stellar lenses.
As we see from Figure 2, the small \shdn\ ``peak" in the MACHO team's
real data set
is distinguishable from the rest of the distribution of \ev\ durations.

\subsubsection{Is an Isolated Event Really Due to Lensing by a Planet?}

Numerical simulations (Figures 4 and 5), the arguments above, 
and the data from the MACHO team, all indicate
that the observing teams are likely to be able to find 
convincing evidence of \shdn\ \ev s, if there is a population of
low-mass lenses. Will the teams be able to determine, however, that the 
\ev s are due to \pl s, rather than to low-mass MACHOs?
A related question is whether a signal due to low-mass MACHOs could be
obscured due to lensing by ordinary \ps s. \footnote{Had   
a large portion of 
the Galactic Halo had been in the form of \pl-mass compact objects,
then the \pl\ signal certainly could not obscure the MACHO signal. The
data collected to date, however, seem to rule this possibility out.} 
It is difficult to come up with comprehensive answers to such questions,
because the distinction between \pl s and MACHOs is not always clear.
For example, when low-mass objects are discovered to be orbiting a pulsar,
they have been called ``\pl s". On the other hand, non-luminous, stellar remnants
in the Galactic Halo and any \pl-mass companions they may have would likely
be considered as MACHOs. Single, non-luminous \pl-mass objects may be
referred to as MACHOs, but some of these may have been ejected from \ps s
with luminous \cs s.

The most stringent criterion we can use to 
classify a low-mass object as a \pl\ is that
it should be in orbit with a luminous star. 
This definition actually gives us a way to test whether a lensing
\ev\ was due to lensing by a \pl . 

If the \cs\ of
a \ps\ lens is luminous, light from the lensed star will
be blended with it. Let us suppose that we observe
a \shdn\ isolated \ev , and that we can establish the following
facts about it.

\noindent (1)
 Not all of the flux emanating from the
line of sight to the \ev\ is emitted by the lensed star.

\noindent (2) 
At least some of the blending is due 
to light from a star whose position cannot be resolved from
that of the lensed star, even with sub-arcsecond resolution. 

In such a case, given the stellar surface density in the Bulge
and LMC, it is almost certain that the additional baseline flux
emanates from the lens system.  
This would definitively establish that the lens giving rise to
the \shdn\ \ev\ was a \pl . In fact, if the amount of light
emanating from the \ps 's \cs\ can be quantified as a function of wavelength,
we may be able to determine its spectral type. 
The observability of blended \ev s is discussed in \rd\ 1998a,b. 
We find that the majority of \ev s should still be detectable
even when there is blending. 

A test of the mass range of the lens (but not the \pl-lens
hypothesis) could be provided by \fsse . 
When spectroscopic studies of the lensed source
can establish its real physical size, finite-source-size effects
may then allow us to derive the true size of the Einstein ring.
This means that,
if the value of $r,$ the ratio between the radius of the lensed star
and the Einstein radius of the lens, 
can be measured from the data,
 $v_T$ can be determined. Furthermore the degeneracy
 in the lens mass associated with the Paczy\'nski \lc\
 is partially broken, since knowing the value of $R_E,$ and, through
 spectral studies, the value of $D_S,$ allows us to
 infer
 \begin{equation}
 M\, x\, (1-x)=C,
 \end{equation}
 \noindent where $C$ is a constant whose value is measured.
 Given the distribution of likely values of $x$ (from $0.9$ to $0.99$,
 for example, in the Bulge) this generally
 constrains $M$ to within a factor of $\sim 5.$ 
Thus, the
 hypothesis that any given \ev\ is due to lensing by a  planetary-mass
 can be meaningfully tested.
Furthermore, if the spectral type of the \cs\ of the lens system
has been determined through the study of blending effects, then
$x$ can be completely determined.

\section{Detection Strategies}

Because good photometric sensitivity increases the rate of observable \ev s and
lengthens the duration of \ev s that are observed, 
the 
the bottom line is 
that 
good photometry is the key to increasing the detection rate of \pl s
in \wo s. Monitoring that is consistently more frequent
than that presently  carried out by the \mo\ teams is
also important, although hourly \mo\ may not be needed,
especially if \fsse\ increase the time during which the
\ev\ is observable.

There is another important point as well. This is that 
the identification of evidence of blending and/or of
\fsse\ can be important. 
Because they can  
possibly clinch the \pl-\ml\ interpretation, it is worthwhile to 
search for evidence of these effects in every \ev . The analysis
of light curve shape
can play an important role; this analysis 
can be done after the \ev\ has ceased.
During the \ev , it would be valuable to obtain spectra,
particularly near the time of peak \mage.

\subsection{Useful Modifications}

The strategies best suited to the discovery of wide planetary systems have
much in common with those best suited to the discovery of planets in resonant
orbits.  In both cases it is important to have frequent time sampling of
events or event features whose duration may be on the order of hours.  In both
cases, good photometry is important if we are to carry out meaningful tests of
the planetary-system lens hypothesis.  Thus, since a detailed search mechanism
is already in place to discover planets in the \zres, it is useful to ask how
effective that mechanism will be at discovering wide planets, and whether
slight modifications of it will make it more effective. 
(See Sackett [1997] for an overview of detection strategies designed
for \pl s in the \zres.)

\subsubsection{Useful Modifications: The Monitoring Teams}

(1) {\sl Spike Analysis:} \
The MACHO and EROS
teams have been able to place upper limits on the numbers of such
\ev s along the direction of the LMC (Renault {\it et al.} 1997, 
Ansari {it et al.} 1996; Alcock {\it et al.} 1996).  
Along the direction to the Bulge it seems likely
that the large majority of the \ev s already
detected are due to lensing by ordinary stars. If a significant
number of stars have planets in \wo s, then short \ev s due
to lensing by these \pl s are expected.
A spike analysis of existing data from the Bulge could be
productive in either discovering evidence of short \ev s 
(in addition to those already discovered) or placing upper
limits on the number of short-time-scale \ev s that might have occurred.
As more data is collected by a larger number of \mo\ teams, systematic spike
analyses should become standard.

(2) {\sl Frequent Monitoring of Some Fields:}\   
The importance of the quest for short time-scale
\ev s suggests that it might be worthwhile for the teams to
each choose one or two fields which they attempt to observe two times
per night. This would allow them to call alerts relatively early
for Jupiter-mass planets, and to have a better chance of finding
evidence of Earth-mass \pl s. In addition, cooperation among the teams
could greatly enhance the probability of finding reliable evidence of
short events. At a recent workshop on \ml , people associated with three
of the teams
discussed the advantages of choosing one or two fields that all of the
teams would attempt to monitor each night. A motivating factor for
such cooperation is to better understand the relative detection
efficiencies of the teams.  The quest for short \ev s provides another
important motivation. Indeed, if each team visited one or two Bulge
fields twice per night, there would at least occasionally, be
6-8 times per night that those fields were checked. If, on average,
10 \ev s per year were to be discovered in those fields,
then, over the course of 3 years, short \ev s could be discovered or
ruled out at the $\sim 3\%$ level.
In addition, the teams should call alerts for \shdn\ \ev s,
even if the \ev\ has apparently ceased before they can announce
its discovery. If the \shdn\ \ev\ is due
to lensing by a \pl, calling the alert 
will allow the follow-up teams to have a better chance of
detecting any subsequent repeat that might be due to lensing by
another object  
in the \ps .
 
(3) {\sl Use of Pixel Techniques:}\ During the past three years, pixel
techniques have begun to be used for the study of \ml\
in M31 (Tomaney \& Crotts 1996; Crotts \& Tomaney 1996; Crotts 1996;
Ansari {\it et al.} 1997; Han 1996).  It has
been estimated that, were such techniques to be
applied to the LMC and Bulge fields, the rate of event detection
would increase by a factor of $\sim 2-3$ (Crotts 1997; Kaplan 1997). 
The reason for the
increase is that observable \ev s can occur when the 
baseline 
flux we receive from a star is not 
bright enough for the star to appear on the templates
presently used by the teams, if the flux
is brought above the detection limits
through \ml. Such \ev s are presently missed by the \mo\ teams.
This increase in detection efficiency, would be helpful, particularly
in any fields singled out for frequent \mo . Indeed, the MACHO team
is presently engaged in applying pixel subtraction techniques to 
their LMC data in an attempt to discover \ev s that were missed by
their standard methods of detection. Application of these techniques to
the
Bulge fields would be also helpful.

\subsubsection{Useful Modifications: The ``Follow-Up" Teams}

While monitoring known ongoing events, the follow-up teams have many
other stars in their field of view.  When the total number, including those
not individually above the detection limit (but which could be brought above
the limit if magnified by some reasonable amount) is large enough, the
follow-up teams can hope to identify new events.  Such observations would
play a unique role in the identification of new events, particularly the
short-duration events that should be associated with wide \ps s.
Programs that would allow the so-called follow-up teams to take the
lead in \ev\ detection are already underway or are planned 
(Sahu 1997; Gould 1997). Indeed,  
an ideal \ml\ search for the purposes of
the detection of planet lenses, is one in which the follow-up
teams play the role now played by the \mo\ teams to discover
\ev s, and continue the work by \mo\ the \lc s of the \ev s they discover.

\section{Conclusions}

Until now, the search for \pl s via \ml\ has focused on a very special 
detection channel. The work presented here and in the companion paper (\rd\ \&
Scalzo 1998) strongly argues for an extension of the search to include
\pl s in \wo s. In this paper we have considered the simplest 
signal of the presence of  
\pl s in \wo s: isolated \ev s of \shdn, that accompany the passage of the track of a distant
star in front of the Einstein ring of an intervening \pl .
Isolated \ev s of \shdn\ may be the ``vanilla" flavor of \pl-lens \ev s. 
They are nevertheless, interesting for several reasons.

(1) If \ps s such as our own are common, then, with
modest changes in detection strategy, we should discover
a few  isolated \ev s of 
\shdn\ 
every year. It seems  unlikely that our own Solar System is an example
of a universal \ps . It is therefore important to note that
a significant fraction of the confirmed \pl\
or brown dwarf systems would also give rise to several, or even
tens of isolated \shdn\ \ev s per year, if {\it they} were typical of a
large fraction of stellar systems. 

(2) Isolated \ev s of \shdn\ may be the most numerous \pl-lens
\ev s. Even though the detection probability for a \pl\ located in the 
\zres\ is enhanced (e.g., for a Jupiter-mass \pl, 
the rate may be as high as $20\%$), there may be a large enough number
of \w-o\ \pl s, for every \pl\ in the \zres, to make the 
detection of isolated \ev s due to \pl s in \wo s the dominant mode
of \pl\ detection.This is particularly so if (a) sensitive photometry can
be employed by the \mo\ teams, and (b) \fsse\ are important.

(3) Searches can be optimized for the detection of isolated \pl-lens
\ev s by implementing sensitive photometry. 
Ideal \mo\ programs would use photometry as sensitive
as that of the present-day follow-up teams, but would also use
difference techniques, in addition to a template, so that they
would be sensitive to the lensing of stars too dim to be on the 
templates. 
Although frequent monitoring is desirable, hourly monitoring
may not be necessary.

(4) Optimized searches can increase the discovery rate significantly.
Event rates for systems like our own Solar System, for example, can be   
doubled. Tests of the hypothesis that all stars are accompanied by
a low-mass (Uranus-mass or larger) companion are well within reach
and can be carried out over the next few years. Less conservative hypotheses,
say that every star is accompanied by a power-of-2 model with 
Saturn-mass companions, can be ruled out or confirmed in a short time.

(5) If all we know from the data sets is that an ensemble
of \shdn\ \ev s has been observed, then we must rely on statistical
arguments (which use the number of stellar-lens \ev s and the numbers
of other types of planet-lens \ev s) to determine what fraction of the \shdn\ \ev s 
may be due to lensing by \pl s. Ideally, however, the analysis of individual
\ev s could tell us some of the following.
(a) the lens is a \pl\ orbiting a luminous star; (b) the spectral type and
mass of the star; (c) the mass of the \pl ; (d) the minimum possible
 orbital separation between the star and \pl . We have pointed out that all of
these determinations may be possible for a subset of \shdn\ \ev s due to \pl s.
This is discussed in more detail elsewhere (\rd\ 1998a,b).

(6) Phenomenologically,
overlap events appear to be isolated \ev s, since the associated \lc\
exhibits just a single deviation from the baseline. Their
\lc s are distinctive,
and provide information about the orbital separation and mass ratio.
Overlap events are long-lived relative to isolated \ev s in which just a single
\pl\ serves as a lens, and thus they are also relatively easy to discover.
The rate of \ov\ \ev s is determined largely by those \pl s
whose physical 
location is within a few Einstein radii of the \cs. For the Solar System, 
for example, Jupiter and Saturn are the most likely to contribute
to the rate of \ov\ \ev s. For reasons of dynamical stability,
there cannot be a large number of such \pl s. The ratio of the
rate of \ov\ \ev s relative to the rate of isolated \ev s of \shdn,  
tells us about the radial structure of \ps s. In particular it
quantifies the size of the
population within several $R_E$ of the \cs , relative to the population
of \pl s in wider orbits. This is illustrated by the comparison between
the power-of-$n$ models, in which planets populate the system
out to $\sim 10^{17}$ cm, and for which the ratio is $\sim 0.1,$
and the Solar System, where the ratio is close to unity.

Although they may be the most frequent \pl -lens \ev s,
isolated \ev s of \shdn\ are
only part of the \pl-lens story. These events are necessarily
accompanied by caustic crossing and \rpe s. The statistics
and characteristics of \ps s
in other galaxies, and in our own Galactic Bulge, can best be 
determined by studying the full complement of \ev s.
In the accompanying paper, we turn our attention to
\rpe s.

\bigskip
\bigskip
\bigskip
\centerline{ACKNOWLEDGEMENTS}

We would like to thank the referee, Scott Gaudi, for suggested changes to
our original manuscript (\rd\ \& Scalzo 1997); his suggestions
have helped to 
improve the presentation. 
One of us (RD) would like to thank Andrew Becker,       
Arlin Crotts, Andrew Gould, Jean Kaplan, Christopher Kochanek, 
David W. Latham, Avi Loeb,   
Shude Mao, Robert W. Noyes, Bodhan Paczy\'nski,
Bill Press, 
Penny Sackett, Kailash Sahu,
Michael M. Shara, Edwin L. Turner,  Michael S. Turner,  
and the participants in the 1997 Aspen workshops, ``The Formation
and Evolution of Planets" and ``Microlensing"
for interesting
discussions, and the Aspen Center for Physics and the Institute for
Theoretical Physics at Santa Barbarba for their hospitality while
this paper was being written,
and the Inter-University Center for Astronomy and Astrophysics in Pune,
India for its hospitality while the paper was revised.
 One of us (RAS) would like to thank the 1996 CfA Summer
Intern Program for support and the Harvard-Smithsonian Center for
Astrophysics for its hospitality while the work was underway.
This work was supported in part by NSF under GER-9450087 and
AST-9619516, and by funding from AXAF.

{}
\end{document}